\documentclass[reprint,
superscriptaddress,
 amsmath,amssymb,
 aps,
 prl,
]{revtex4-1}

\usepackage{graphicx}
\usepackage{dcolumn}
\usepackage{bm}
\usepackage{amsmath}
\usepackage{color}
\usepackage{float}
\usepackage[hidelinks]{hyperref}

\usepackage[section]{placeins}

\graphicspath{{./Figures/}}  

\newcommand{\prlsec}[1]{\textit{#1.---}}

\newenvironment{extraabstract}
{\small
  \begin{center}
    \bfseries \vspace{-.5em}\vspace{0pt} 
  \end{center}
  \quotation}
 {\endquotation}

\begin{document}

\pretolerance=10000
\tolerance=2000
\emergencystretch=10pt

\displaywidowpenalty = 10000

\title{Real-Time Kalman Filter: 
Cooling of an Optically Levitated Nanoparticle }

\author{Ashley Setter}
\email{A.Setter@soton.ac.uk}
\affiliation{%
 Department of Physics and Astronomy, University of Southampton, SO17 1BJ, United Kingdom 
}%

\author{Marko Toro\v{s}}
\affiliation{%
 Department of Physics and Astronomy, University of Southampton, SO17 1BJ, United Kingdom 
}%

\author{Jason F. Ralph}%
\affiliation{%
 Department of Electrical Engineering and Electronics, University of Liverpool, L69 3GJ, United Kingdom 
}%

\author{Hendrik Ulbricht}%
 \email{H.Ulbricht@soton.ac.uk} 
\affiliation{%
 Department of Physics and Astronomy, University of Southampton, SO17 1BJ, United Kingdom 
}%

\begin{abstract}
  We demonstrate that a Kalman filter applied to estimate the position of an optically levitated nanoparticle, and operated in real-time within a Field Programmable Gate Array (FPGA), is sufficient to perform closed-loop parametric feedback cooling of the centre of mass motion to sub-Kelvin temperatures. The translational centre of mass motion along the optical axis of the trapped nanoparticle has been cooled by three orders of magnitude, from a temperature of $300\text{K}$ to a temperature of $162 \pm 15\text{mK}$.
\end{abstract}

\maketitle

\prlsec{Introduction} In order to perform closed-loop feedback control of a classical or quantum system, accurate real-time state estimation is crucial. With recent advances in quantum engineering and technology comes a need to accurately measure and control quantum systems. In order to obtain accurate knowledge about the state of a system from noisy measurements one can use a process called filtering which combines the knowledge of the dynamics of the system with noisy measurements of the system to estimate the true state of the system \cite{Jazwinski1970}.

A much targeted goal in levitated optomechanics \cite{Aspelmeyer2013} is cooling and stabilising the centre of mass motion of an optically levitated nanosphere in a target phonon number state \cite{Chang2010,Romero-Isart2010,Romero-Isart2011,Kiesel2013,Genoni2015,Mestres2015,Jain2016,Vovrosh2017,Barker2017,Ralph2017}. Nanospheres cooled to low temperature thermal states and stabilised in phonon number states have applications such as performing matter-wave interferometry, allowing investigation of quantum phenomena which cannot be accessed with atoms \cite{Bateman2014a}, and tests of collapse models which cannot be performed at smaller mass scales \cite{Hornberger2012,Eibenberger2013,Bera2015}; as well as providing the possibility of much higher force sensitivity than can be achieved in levitated atom systems \cite{Gieseler2013,Jain2016,Monteiro,Hempston2017a}. Optically levitated nanoparticles have also been used as a model system to simulate and investigate non-equilibrium dynamics \cite{Lechner2013} and stochastic dynamics \cite{Ricci2017} and show promise for use in investigating quantum gravity \cite{Albrecht2014,Carlesso2017,Bassi2017}.

The first step in stabilising the centre of mass motion in such a state using closed-loop feedback is to accurately estimate the motional state of the system in real-time. In principle, this can be accomplished using a quantum filter, known also as the stochastic master equation (SME), where the estimate of the state, i.e. the conditional density operator, is updated continuously by the measurement record~\cite{Belavkin,Belavkin1992,Belavkin1992a,Belavkin1994}. Quantum estimation theory has been discussed in the context of optomechanics to investigate wavefunction collapse models \cite{Genoni2016a, McMillen2017}, to detect and measure gravity \cite{Qvarfort2017, Armata2017} and the Fisher information for parameter estimation in linear Gaussian quantum systems under continuous measurement has been considered \cite{Genoni2017}.

A quantum filter is a more general approach to modelling the system than a Kalman-Bucy filter as it includes the effect of measurement backaction on a quantum system under continuous weak measurement. However, as will be shown here, it reduces to the optimal Quantum Kalman-Bucy filter in the case where the system is linear and the noise is approximately white and Gaussian~\cite{Belavkin, doherty2000quantum, Gough2009, Miao2012}. Moreover, as discussed in~\cite{Doherty}, we can formally map the Quantum Kalman-Bucy filter to a (classical) Kalman-Bucy filter~\cite{Kalman1960,Kalman1961}. Kalman filters, the discrete-time counterparts to continuous-time Kalman-Bucy filters, have been used extensively in many aerospace and defence applications~\cite{Blackman,Bar-Shalom}, including navigation systems for the Apollo Project and the well-known Global Positioning System (GPS)~\cite{5466132}. FPGA based Kalman filters have also recently been developed for applications including antilock braking systems~\cite{Sandhu2017}, radar tracking systems~\cite{Kara2017} and displacement measuring interferometry~\cite{Wang2017}. Kalman filtering has also been applied in various areas within the physical sciences such as atomic magnetometry \cite{Geremia2003}, tracking dusty plasmas \cite{Oxtoby} and noise cancellation in gravitational wave detection \cite{SamuelFinn2000}. 

The second step is to control the state of the system with feedback~\cite{wiseman2002bayesian,Dong2010}, e.g. Markovian~\cite{wiseman1993quantum} or Bayesian feedback~\cite{Doherty}, the latter of which we will adopt in this letter (alternatively one could also consider coherent feedback~\cite{wiseman1994all}).  However, numerically solving a SME in real-time requires truncation of the Hilbert space basis as the computation time scales exponentially with the size of basis. To circumvent this difficulty different suboptimal methods have been developed, namely, the number-phase Wigner particle filter~\cite{hush2013controlling}, the Volterra particle filter~\cite{tsang2015volterra}, the quantum extended filter~\cite{emzir2017quantum} and the Gaussian approximation of the conditional density operator~\cite{doherty2000quantum}. 

Quantum Kalman filtering has recently been applied to the field of optomechanics and demonstrated to produce a minimal least-squares estimation of the mechanical state of an optical cavity \cite{Wieczorek}. This was done in post-processing by using the measurement record from a homodyne detection as the input to a Quantum Kalman filter implementing an accurate state-space model which carefully took into account nontrivial experimental noise sources. They suggest that ground state cooling should be readily achievable by utilising a real-time Quantum Kalman filter of sufficiently high spatial resolution, dynamic range and latency in the detection and processing of the signals. 

In this letter we demonstrate that a real-time Quantum Kalman filter with a sampling period of $2.275\mu s$ is sufficient to estimate the state such that one can perform closed-loop parametric feedback cooling of the translational motion of an optically levitated nanoparticle to sub-Kelvin temperatures. 

\begin{figure}[t!]
  \centering
  \includegraphics[width=0.45\textwidth]{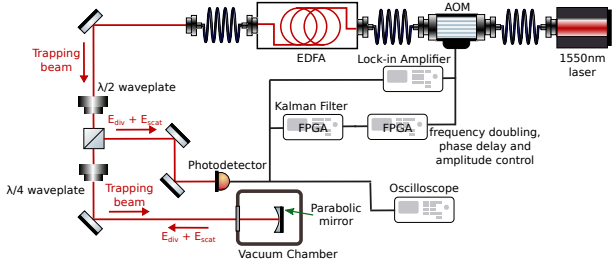}
  \caption{\label{ExperimentalSetup}
    The 3D position of the particle is detected by interference of the scattered and divergent field by the photo-detector. This signal is then passed into an FPGA which is implementing the Kalman filter that estimates the $z$ position and amplifies the estimate. This estimated signal is then sent into a second FPGA which DC shifts it, frequency doubles it, applies a time delay and multiplies it such as to keep the amplitude approximately constant. This signal is then fed to a AOM to modulate the power of the trapping lasers. The Erbium Doped Fibre Amplifier (EDFA) then amplifies this modulated signal to the power required to trap the nanoparticle, this signal is then used to trap and measure the nanoparticle.
  }
\end{figure}

\prlsec{The model}
We consider an optically levitated nanoparticle in free space subject to continuous weak measurements of its position. Specifically, we consider the experimental setup described in~\cite{rashid2017wigner}: an incoming beam is focused by a paraboloidal mirror to the focal point, where it creates a harmonic trap. The particle, which is trapped in this potential scatters photons in the  Rayleigh regime: these are collected by the detector to obtain the $z$ position with efficiency $\eta$ (see Fig.~\ref{ExperimentalSetup}). We model the particle dynamics along the optical axis, namely the $z$-axis, using the following SME:
\begin{equation}
\begin{split}
d\hat{\rho_c}=&-\frac{i}{\hbar}[\hat{H}+\hat{H}_\text{fb},\hat{\rho}_\text{c}]dt\\
&+(\bar{n}+1)\Gamma \mathcal{D}[\hat{a}]\hat{\rho}_c dt
+\bar{n}\Gamma \mathcal{D}[\hat{a}^{\dagger}]\hat{\rho}_c dt \\
&+2 k \mathcal{D}[\hat{z}]\hat{\rho}_c dt+\sqrt{2\eta k}\mathcal{H}[\hat{z}]\hat{\rho}_c dW,
\end{split}
\label{sme}
\end{equation}
where $\hat{\rho}_c$ is the conditioned state at time $t$. On the first line we have Hamiltonian and feedback terms:
\begin{align}
\hat{H}&=\frac{\hat{p}^2}{2m}+\frac{m\omega^2}{2}\hat{z}^2,\\
\hat{H}_\text{fb}&=\beta\left( \frac{m\omega^3}{2}
\frac{   \langle \hat{p}\rangle  \langle \hat{z}\rangle}{\langle \hat{H}\rangle}
\right) \hat{z}^2,
\end{align}
respectively, where $\beta$ is an adimensional parameter quantifying the strength of the feedback, $\langle \, \cdot  \, \rangle=\text{tr}[\, \cdot  \,\rho_c]$, $\hat{z}$ and $\hat{p}$ denote the particle position and momentum operators, respectively, $\omega$ is the trap frequency and $m$ is the mass of the particle (see appendix A1 for more details on the feedback term). The second line of Eq.~\eqref{sme} describes the interaction with a gas of particles at temperature $T$, where $\bar{n}=(\text{exp}(\hbar\omega / k_B T)-1)^{-1}$, $k_B$ is Boltzmann's constant, 
$\hat{a}^{\dagger}=\sqrt{\frac{m \omega}{2\hbar}}(\hat{z}+\frac{i}{m\omega}\hat{p})$,
$\mathcal{D}[\hat{L}]\,\cdot\,=\hat{L}\,\cdot\,\hat{L}^\dagger-\frac{1}{2}\{hat{L}^\dagger \hat{L}, \,\cdot\, \}$~\cite{wiseman1993interpretation}, $\hat{L}$ denotes an operator, $\{\,\cdot \, , \,\cdot \,\}$ is the anti-commutator and $\Gamma$ is the damping rate~\cite{spiller1993thermal}. The effect of weak continuous $z$-position measurements is described by the third line of Eq.~\eqref{sme}, where $\mathcal{H}[\hat{L}]\,\cdot\,=\{\hat{L},\,\cdot\,\}-2\text{Tr}[\hat{L}\,\cdot\,]\,\cdot\,$ and $\hat{L}$ is a Hermitian operator, $W$ is a real Wiener process, $k=\frac{12\pi^2\mu}{5\lambda^2}$, $\lambda$ is the wavelength of the laser light, $\mu=\frac{\sigma}{\pi w_0^2}\frac{P}{\hbar \omega_L}$ is the scattering rate, $\sigma$ is the the Rayleigh cross section, $w_0$ is the beam waist, P is the laser power, $\omega_L=\frac{2\pi c}{\lambda}$ and $c$ is the speed of light. In addition, we assume that the measurement record is given by~\cite{rashid2017wigner}:
\begin{equation}
d Q=4 \eta k \langle \hat{z}\rangle dt+\sqrt{2\eta k}dW.
\label{record}
\end{equation}

We suppose that the initial state $\rho_c$ is thermal, when the feedback term starts to cool the system, and thus the state $\rho_c$ remains Gaussian under the evolution of the SME given in Eq.~\eqref{sme}. This simplifies the problem to the analysis of the mean values~\cite{doherty2000quantum}:
\begin{align}
d \langle \hat{z} \rangle =& \frac{\langle \hat{p} \rangle}{m}dt-
\Gamma\langle \hat{z} \rangle dt
+\sqrt{8 \eta k}V_z dW, \label{zeq}\\
d\langle \hat{p} \rangle =&  -m\omega^2\langle \hat{z} \rangle 
\left(1+ \beta \frac{\omega \langle \hat{p}\rangle  \langle \hat{z}\rangle}{\langle \hat{H}\rangle} \right)dt \nonumber\\
&-\Gamma\langle \hat{p} \rangle dt
 +\sqrt{8 \eta k} C dW, \label{peq}
\end{align}
and of the covariances:
\begin{align}
d V_z = & \frac{2}{m} C dt -8\eta k V_z^2dt\nonumber\\
&-\Gamma V_z dt+\Gamma(2\bar{n}+1)\frac{\hbar}{2m\omega}dt
-3 \Gamma  \langle z\rangle^2 dt, \label{Vz}\\
d V_p = & -2m \omega^2 C \left( 1+\beta  \frac{\omega \langle \hat{p}\rangle  \langle \hat{z}\rangle}{\langle \hat{H}\rangle}\right)dt-8\eta k C^2 dt
+ 2 k\hbar^2 dt\nonumber\\
&-\Gamma V_p dt 
+\Gamma(2\bar{n}+1)\frac{m\omega\hbar}{2} dt -3\Gamma \langle \hat{p}\rangle^2 dt ,\label{Vp}\\
d C = &  \frac{V_p}{m} -m\omega^2 V_z\left( 1+\beta  \frac{\omega \langle \hat{p}\rangle  \langle \hat{z}\rangle}{\langle \hat{H}\rangle}\right)dt-8\eta k C V_zdt\nonumber\\
&- \Gamma Cdt - 3 \Gamma \langle \hat{p}\rangle  \langle \hat{z}\rangle dt, \label{C}
\end{align}
where
$V_z=\langle (\hat{z} - \langle \hat{z} \rangle )^2\rangle$, 
$V_p=\langle (\hat{p} -\langle \hat{p}  \rangle)^2\rangle$ and $C=\frac{1}{2}\langle \{\hat{z},\hat{p} \}\rangle-\langle \hat{z} \rangle \langle \hat{p} \rangle$.

We can further simplify the filter by neglecting the small feedback term and damping term, i.e. we set $\beta=0$ and $\Gamma=0$ in Eqs.~\eqref{zeq}-\eqref{C}. The equations for the variances then form a closed set of coupled Riccati equations~\cite{belavkin1989lecture,BELAVKIN1989359,
barchielli1993stochastic,chruscinski1992asymptotic} and we can also formally rewrite Eqs.~\eqref{zeq} -~\eqref{C} as a classical Kalman-Bucy filter~\cite{Doherty}: 
\begin{equation}
d \bm{x}_c = A \bm{x}_c dt+\sqrt{2 \eta k} d\bm{\xi}+ \sqrt{2\Gamma k_b T} d\bm{V}, \label{kb}
\end{equation}
where $\bm{x}_c=(z_c,p_c)^\top$, 
$A=\begin{bmatrix}
  0 & 1/m \\
  -m\omega^2 & -\Gamma
 \end{bmatrix}$,
$d\bm{\xi}=(0,d\xi)^\top$, $\xi$ is a real Wiener process,
$d\bm{V}=(0,dV)^\top$, $V$ is a real Wiener process uncorrelated with $\xi$, and we have added the $\Gamma$ dependant terms on $p_c$. In place of Eq.~\eqref{record} we consider the classical measurement record:
\begin{equation}
dQ_c=4\eta k z_c dt+\sqrt{2\eta k}d\zeta,
\end{equation}
where $d\zeta$ is a third real Wiener process uncorrelated with $\xi$ and $V$. Moreover, we suppose the following relation:
\begin{equation}
dW=\sqrt{8\eta k} (z_c-\mathbb{E}[z_c])dt+d\zeta,
\end{equation}
where $\mathbb{E}[ \,\cdot\,]=\int \,\cdot\,P_c(z_c,p_c) dz_c dp_c$ and $P_c$ is the (classical) conditioned state obtained from the Kushner-Stratonovich equation corresponding to Eq.~\eqref{kb}~\cite{Wiseman2010}. We can then formally identify $\mathbb{E}[\, O \,]$ with $\langle \hat{O}\rangle$, where $O$ and $\hat{O}$ denote the classical and its corresponding quantum observable, and $Q_c$ with $Q$~\footnote{In the classical Kalman-Bucy filter we have neglected the position damping term $-\Gamma x_c$. The noise $\xi$ models the environment at temperature $T$, which damps only the momentum $p_c$.}.

\prlsec{Experimental Methods} The Kalman filter described in Eq.~\eqref{kb} and further detailed in appendix (2) was implemented in VHDL (Very High Speed Hardware Description Language) using fixed-point arithmetic and synthesised onto a XilinX Virtex-5 SX50T Field Programmable Gate Array (FPGA) provided in a National Instruments (NI) PXIe-7961 and connected to a NI 5781 baseband transceiver for ADC and DAC conversion. The fastest sample rate achievable for a Kalman filter with this FPGA was 439.56kHz, this was because the fastest synthesisable clock rate for the design was 3.07MHz and each Kalman filter iteration takes 7 clock cycles, this is equivalent to a sample period of $2.275\mu s$.

Cooling was performed by taking the estimated signal for the $z$ position from the Kalman filter and using a leaky integrator to calculate the DC component of the signal. The DC component was then removed from the measured signal in order to provide only the AC component of the signal. This AC signal containing the estimate of the particle position was squared in order to produce a signal at double the frequency of the motion of the particle. The signal then had a constant phase offset applied to it by introducing a time delay, in order to compensate for experimental latency, which was optimised such that maximal cooling was observed experimentally. In addition, the amplitude of the output cooling signal was controlled such that it maintained a set average amplitude in order to keep the cooling rate approximately constant regardless of fluctuations in the amplitude of the motion of the particle. This signal was then applied to modulate the power of the trapping laser in order to parametrically cool the translational $z$ motion.

A particle was trapped and the power of the trap was lowered such as to reduce the frequency of oscillation in the $z$ direction, a frequency of 38kHz was obtained for the $z$ motion. The VHDL code for a Kalman filter modelling a simple harmonic oscillator with a frequency of 38kHz was generated, the $\mathbf{Q}$ matrix and $R$ value, corresponding to Eq. (\ref{StateOfSystem}) and Eq. (\ref{MeasurementOfSystem}) in appendix 2, were tuned by application to simulated data \footnote{Simulated data was generated by solving the classical stochastic differential equation modelling a levitated nanosphere using the optosim package \cite{Setter2017}} and then synthesised onto an FPGA. The frequency doubling, time delay and amplitude control code was synthesised onto another FPGA. The experimental setup is shown in figure \ref{ExperimentalSetup}.

A lock-in amplifier was used to cool the other two directions of motion as described in reference \cite{Vovrosh2017}. The power of the trapping laser was adjusted slightly so that the $z$ motion, the motion parallel to the propagation direction of the laser, stayed oscillating at a frequency of 38kHz regardless of pressure so that the Kalman filter could continue to optimally track the particles $z$ motion.

\begin{figure}[htp]
  \centering
  \includegraphics[width=0.5\textwidth]{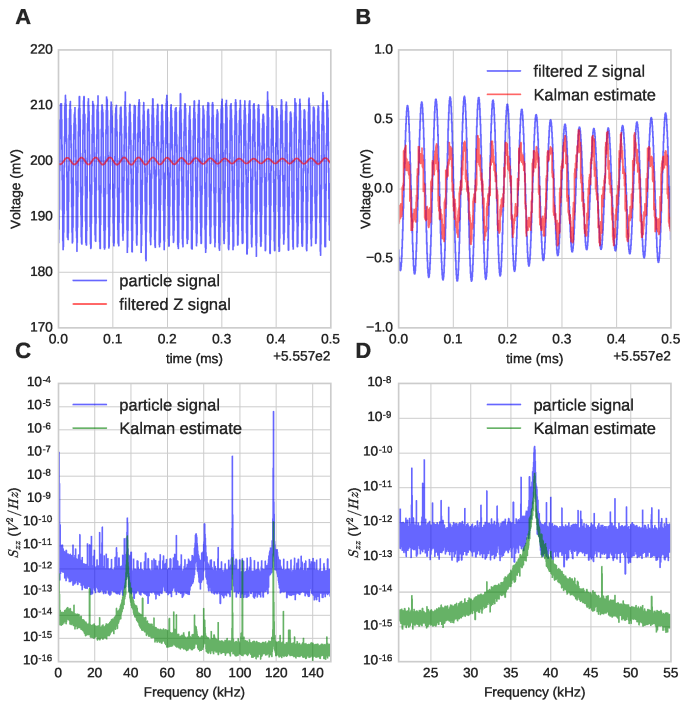}
  \caption{\label{KalmanTracking}
    (A) Time trace of unfiltered measured signal from particle and filtered $z$ signal over 0.5ms. (B) Time trace of filtered $z$ signal and Kalman filter estimated $z$ signal over 0.5ms. (C) Power spectral density of measured signal from particle and the Kalman filtered estimate, the peaks at 100kHz and 120kHz are due to the $x$ and $y$ motion respectively. (D) Same as C but over a smaller frequency range centred on the $z$ frequency of 38kHz.
  }
\end{figure}

\prlsec{Results} The data analysis has been primarily performed using the open source optoanalysis package which we have developed \cite{Setter2017b}. The derivation of the method of calculating temperature is detailed in reference \cite{Vovrosh2017}. Plots showing the measured, bandpass filtered and real-time Kalman estimated $z$ motion signal are shown in figure \ref{KalmanTracking}. The Kalman estimate has been shifted forward in time by one filter cycle (time for one time-step / iteration of the Kalman filter algorithm which is $2.275\mu s$) to account for the latency in the estimation. The dominant noise source in the estimated and cooling signals is electrical noise originating from the Analogue to Digital and Digital to Analogue Converters.

\begin{figure}[t!]
  \centering
  \includegraphics[width=0.5\textwidth]{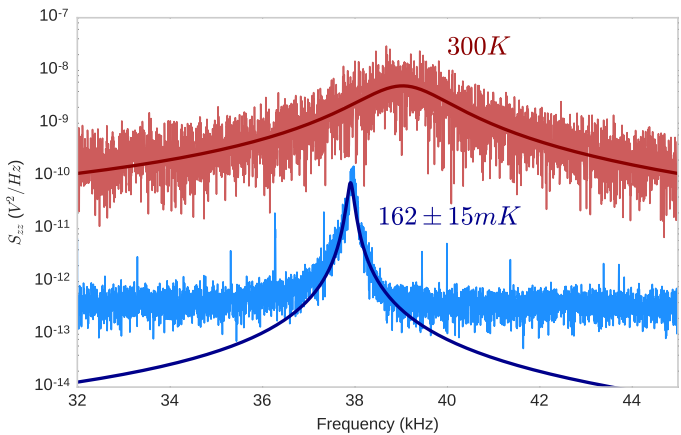}
  \caption{\label{KalmanCoolingResult}
    The PSD of the uncooled particle in equilibrium with the environment at 300K (at a pressure of 3mbar) and the cooled particle at a pressure of $5.7\times 10^{-5}$ mbar. The lines represent the Lorentzian functions fitted to the PSD data to calculate the temperature of the cooled state.
  }
\end{figure}

As described in the Experimental Methods section this estimated signal was then passed to a second FPGA so that the signal could be squared, a time delay applied and a modulation applied to the amplification of the signal so as to keep the amplitude approximately constant. This signal was then used to modulate the power of the laser via an Acousto-Optic Modulator (AOM) as shown in figure \ref{ExperimentalSetup}. Through this technique cooling of the translational motion in the propagation direction of the laser (labelled the $z$ direction) from $300$K to $162 \pm 15$mK was achieved, see figure \ref{KalmanCoolingResult} for the PSD (power spectral density) of the uncooled and cooled signal.

The main limitations on the temperature reached with this cooling scheme are discretisation noise from the ADC, stochastic noise on the particle motion from gas collisions and the sample period with which one can perform an iteration of the Kalman filter state estimation. 

The discretisation noise is caused because the voltage signal from the photodetector is read into the FPGA as a digital value from an ADC. For the ADC used here the voltage difference between the discrete observable levels is 122$\mu$V, this means that for the signal shown in figure \ref{KalmanTracking} the FPGA only observed $\sim5$ discrete levels. The feedback cooling is operating very near the limit of what motion it can discern and estimate, and this is likely the predominant limiting factor on the cooling achievable. Using an ADC with higher voltage resolution or amplifying the signal into the ADC with a sufficiently low noise amplifier will improve this limit.

The stochastic noise on the motion of the particle from gas collisions is the other primary limitation on the temperature reached; performing this cooling at lower pressures will further reduce the temperature that can be reached as the stochastic driving of the particle motion by gas collisions will be further reduced.

At present the hardware implementation can run at a fastest sample period of $2.275\mu s$, this means that for a particle oscillating with a frequency of 38kHz one only estimates the motion $\sim11.6$ times in one period and results in a higher signal to noise ratio being necessary for accurate state estimation. Increasing the sample frequency would mean that the accuracy of the state estimation at lower signal to noise ratios will improve and therefore that cooling to lower temperatures should be achieved. A high sample frequency also means that higher frequency motion can be estimated and cooled in this way, leading to a lower mean phonon number for the same temperature of motion \cite{Gieseler2012,Vovrosh2017}.

The model used in this Kalman filter is sufficient to out-perform other, more simplified, state estimation and feedback cooling schemes such as using a bandpass filter to track and cool the motion, which we found only achieved cooling to a temperature of $3K$. However, the model of the motion considered here has been simplified to that of a simple harmonic oscillator which does not represent a perfect model of the experimental system, because of this the state estimation is suboptimal which reduces the accuracy with which the Kalman filter can estimate the motional state of the system, which in turn causes the cooling to be suboptimal. A more sophisticated model that more completely modelled the physical system would produce a more accurate estimate and therefore a higher cooling rate which would result in an achievable final temperature which was lower. 

\prlsec{Discussion and Conclusions}
We have demonstrated that, for an optically levitated nanoparticle, a Kalman filter using a very simple harmonic-oscillator model of the dynamics of the system and which operates with a relatively high sample period is sufficient to achieve cooling of translational motion to sub-Kelvin temperatures of $162\pm15$mK. Improvements in the speed of the hardware implementation such that a lower sample period can be achieved, a more sophisticated model of the dynamics including the effect of feedback and extending the modelling to all 3 translational degrees of freedom, as well as performing the cooling at a lower pressure, will increase the performance of the cooling performed using this method and result in cooling to lower temperatures. Provided that the limitations discussed could be overcome and a sufficiently accurate model of the system could be implemented in a Kalman filter and that the achievable cooling rate could be made sufficiently high to overcome the rate of re-heating caused by thermal gas heating and photon recoil then ground state cooling should be achievable in this way.

Using this form of state estimation in real-time also opens the way to implementing more complex feedback schemes \cite{Wiseman2010,Zhang,Jacobs2006,Jacobsa}, such as combination with a Proportional Integral Differential (PID) controller \cite{Gough} or a linear quadratic regulator (LQR)~\cite{Doherty,Wiseman2010}. Combining the Kalman filter, which is a linear quadratic estimator (LQE), with an LQR constitutes linear quadratic Gaussian (LQG) feedback control. Coherent-feedback cooling by LQG Control has been discussed in depth in~\cite{Hamerly2012} and \cite{Hamerly2013} and could be used to significantly improve the level of control over the motional state of the system as well as improving the minimum temperature reachable in this system.

\prlsec{Acknowledgements}
We would like to thank  M. Bilardello, M. Rashid, C. Timberlake, L. Ferialdi and J. Murrel for discussions. We also wish to thank the Leverhulme Trust and the Foundational Questions Institute (FQXi) for funding. A. Setter is supported by the Engineering and Physical Sciences Research Council (EPSRC) under Centre for Doctoral Training grant EP/L015382/1. We also acknowledge support from EU FET project TEQ (grant agreement 766900). All data supporting this study are openly available from the University of Southampton repository at https://doi.org/10.5258/SOTON/D0355. The code used to analyse the data is openly available at https://doi.org/10.5281/zenodo.1042526.

\bibliography{library.bib} 

\onecolumngrid
\newpage

\begin{center}
  \part{\centering \large Appendix for: \\ Real-Time Kalman Filter: Cooling of an Optically Levitated Nanoparticle' }
\end{center}
\begin{extraabstract}
  \small In this appendix we derive the form of the feedback term in the stochastic master equation (SME) in Eq.~\eqref{sme} in the main text. We also derive the discrete-time state transition matrix for the Kalman-Bucy filter considered in Eq.~\eqref{kb} in the main text.
\end{extraabstract}

\section*{A1: The feedback term}

We denote the measured signal by $\bar{z}=Q$, where $Q$ is given in Eq.~(4) in the main text, and we define the phase of the signal as:
\begin{equation}
  \theta_t= 
\begin{cases}
    \text{arctan}\left(\frac{\bar{p}}{m \omega \bar{z}}\right),& \text{if } \bar{z}\geq 0,\\
    \text{arctan}\left(\frac{\bar{p} }{m \omega \bar{z}}\right)+ \pi,              & \text{otherwise},
\end{cases}
\label{theta}
\end{equation}
where $\bar{p}=m\frac{d}{dt}\bar{z}$. Note that this definition of $\bar{p}$ assumes a physical noise with a non-white spectrum in place of the Brownian noise $\frac{dW}{dt}$. We apply a sinusoidal modulation of the laser power $P$ at twice the tracked signal phase. Specifically, we consider the modulation $\beta\text{sin}(2\theta_t)$, where $\beta$ is the amplitude of the modulation of the laser power $P$. We have that the trap frequency squared $\omega^2$  is proportional to laser power $P$ (see main text and~\cite{rashid2017wigner}). Thus when the modulation is applied we can obtain the feedback term by formally making the replacement $\omega^2 \rightarrow \omega^2(1+\beta\text{sin}(2\theta_t))$ in the unmodulated Hamiltonian $H$ (we neglect the corresponding modulation of the $k$ dependant terms in Eq. \eqref{sme}). The feedback term is given by:
\begin{equation}
H_\text{fb}=\frac{\beta m\omega^2}{2}\text{sin}(2\theta_t) \hat{z}^2.
\label{hfb}
\end{equation}
Using the definition in Eq.~\eqref{theta} we note that:
\begin{equation}
\text{sin}(2\theta_t)=\text{sin}\left(2\arctan \left(\frac{\bar{p} }{m \omega \bar{z}}\right)\right).
\label{sin2th}
\end{equation}
Using Eq.~\eqref{sin2th} and trigonometric identities we then find from Eq.~\eqref{hfb}:
\begin{equation}
H_\text{fb}=\beta \left(\frac{m\omega^3}{2} 
\frac{ \bar{p}\bar{z}}{\bar{E}}
\right) \hat{z}^2,
\label{hfb1}
\end{equation}
where $\bar{E}=\frac{\bar{p}^2 }{2m}+\frac{m\omega^2}{2}\bar{z}^2$. Note that $\bar{E}$ is also time dependent. 

It is a non-trivial task to add the feedback term in Eq.~\eqref{hfb1} to the dynamics in Eq.~(1) due to the non-white nature of the noise. One would need to find the white noise limit of the term in Eq.~\eqref{hfb1}, for example for a Gaussian noise; add it to the dynamics in the Stratonovich form, and then convert back to the It\^{o} form~\cite{wiseman1993quantum}. However, if one considers only the non-stochastic contribution, then the feedback term in Eq.~\eqref{hfb1} reduces to:
\begin{equation}
H_\text{fb}=\beta \left(\frac{m\omega^3}{2} 
\frac{ \langle \hat{p} \rangle \langle \hat{z}\rangle}{\langle \hat{H} \rangle}
\right) \hat{z}^2.
\end{equation}
This term yields, in the equations of motion for $\langle \hat{p} \rangle$, the following contribution:
\begin{equation}
d \langle \hat{p} \rangle = \left(
\frac{\beta m\omega^3 }{\langle \hat{H} \rangle}  
\right) \langle \hat{z}\rangle^2 \langle \hat{p} \rangle dt,
\end{equation}
If we then replace $\langle \hat{H} \rangle$ with a constant energy value $E$ we obtain the cooling term considered in~\cite{gieseler2015non,Vovrosh2017}.

\section*{A2: The Discrete-time Kalman filter}

The discrete-time Kalman filter uses a linear state space model of the form

\begin{equation} \label{StateOfSystem}
\mathbf{X}_t = \bm{F_t} \mathbf{X}_{t-1} + \mathbf{w}_t \text{,}
\end{equation}
where $\mathbf{X}_t$ is the state vector containing the variables that one wishes to estimate (e.g. position, velocity), $\bm{F_t}$ is the state transition matrix which describes how the state vector at time step $t-1$ transitions to the state vector at time step $t$, $\mathbf{w}_t$ is the vector containing the discrete process noise for each parameter in the state vector. The process noise is assumed to be drawn from a zero mean multivariate normal distribution with covariance matrix $\bm{Q_t}$ i.e. $\mathbf{w}_t \sim \mathcal{N}(0, \bm{Q_t})$.
Measurements of the system take the form

\begin{equation} \label{MeasurementOfSystem}
\mathrm{\mathbf{z}}_t = \bm{H}_t \mathbf{X}_t + \mathbf{v}_t \text{,}
\end{equation}
where ${\mathrm{\mathbf{z}}}_t$ is the vector of measurements, $\bm{H_t}$ is the measurement transformation matrix which maps the state vector domain to the measurement domain, $\mathbf{v}_t$ is the vector containing the measurement noise terms for each observation in the measurement vector. The measurement noise, similar to the process noise, is assumed to be drawn from a zero mean normal distribution with covariance $R_t$, i.e. $\mathbf{v}_t \sim \mathcal{N}(0, R_t)$.

If one considers the Kalman-Bucy filter given in Eq. (10) of the main text with the state vector $\mathbf{X_t} = (z_t, v_t)^T$ at time step $t$, where $z_t$ is the position of the particle in the $z$ direction and $v_t = p_t/m$ is the velocity of the particle in the $z$ direction, one has the following equation for the system dynamics, 

\begin{equation}
  \mathbf{\dot{X}}_t = \mathbf{A} \mathbf{X}_t + \bm{\omega} \text{,}\\
\end{equation}
where 
\begin{align}
\mathbf{A} =& 
\begin{bmatrix}
  0 & 1\\
  -\omega^2 & -\Gamma \\
\end{bmatrix}, 
\,\,\,\,\,\,
 \bm{\omega} = 
  \sqrt{\dfrac{2\Gamma k_B T_0}{m}} \dfrac{d\bm{\xi}(t)}{dt} + \sqrt{\dfrac{2\eta k}{m}} \dfrac{d\mathbf{V}(t)}{dt},
\end{align}
$d\bm{\xi}=(0,d\xi)^\top$ and   $d\bm{V}=(0,dV)^\top$.

The stochastic terms can is modelled as process noise $\bm{\omega}$. The damping is time variant as it varies with pressure, however for the range of pressures explored experimentally $\Gamma << \omega^2$ and therefore the damping value can be approximated as 0. In this case the dynamics model simplifies to a simple sinusoidal model of the motion $\ddot{x} = -\omega^2 x$ with the stochastic noise modelled in the process noise. This simple model also allows us to keep the hardware implementation simple to allow the design to fit inside the FPGA. 

As described in references \cite{Schwarz1965,Labbe, Musoff2005} we can use the following transformation from linear time invariant system theory (LTI System theory) in order to calculate the continuous-time state transition matrix $\mathbf{F}(t)$ for a time-invariant systems dynamics matrix $A$,

\begin{equation} \label{LTI_Transform}
  \mathbf{F}(t) = \mathcal{L}^{-1}([s \mathbf{I} - \mathbf{A}]^{-1}) \text{,}
\end{equation}
where $\mathbf{F(t)}$ is the continuous-time form of the $\mathbf{F_t}$ matrix in Eq.~\eqref{StateOfSystem}, $\mathcal{L}^{-1}$ is the inverse Laplace transform in terms of the complex frequency variable $s$ and $\mathbf{I}$ is the identity matrix. Performing this transformation and taking continuous time $t$ to be in discrete steps $\Delta t$, gives us the discrete time state-space model
      
\begin{equation} 
\begin{bmatrix}
  z_t \\
  v_t \\
\end{bmatrix}
=
\left[\begin{matrix}\cos{\left (\omega \Delta t \right )} & \frac{1}{\omega} \sin{\left (\omega \Delta t \right )}\\- \omega \sin{\left (\omega \Delta t \right )} & \cos{\left (\omega \Delta t \right )}\end{matrix}\right]
\begin{bmatrix}
  z_{t-1} \\
  v_{t-1} \\
\end{bmatrix}
  \text{,}
\end{equation}
where $\Delta t = t_{n} - t_{n-1}$.

The values of the process noise covariance matrix $\mathbf{Q}$ and the measurement noise covariance value $R$ were tuned such that when the HDL (Hardware Description Language) implementation of the Kalman filter was run on noisy data produced by a simulated signal it produced an accurate estimate of the true signal. This simulated signal was produced by adding Gaussian noise to the simulated position measurements found by solving the classical stochastic differential equation modelling our system under free evolution using the open source optosim package we have developed \cite{Setter2017}.

\end{document}